
\def\sect{\vskip 2mm \centerline}
\def\se{\vskip 2mm \centerline}
\def\r{\hangindent=1pc  \noindent}
\def\ref{\hangindent=1pc  \noindent}
\def\cen{\centerline}
\def\ce{\centerline}
\def\v{\vskip 1mm}

\def\endpage{\vfil\break}
\def\page{\vfil\break}
\def\noi{\noindent}
\def\kms{km s$^{-1}$}

\def\deg{$^\circ$}

\def\Msun{M_{\odot \hskip-5.2pt \bullet}}

\def\MH{m_{\rm H}}
\def\mH{$m_{\rm H}$}
\def\halpha{H$\alpha$}

\def\pcc{cm$^{-3}$}

\def\deg{$^\circ$}

\def\htwo{H$_2$}

\def\pa{PASJ}
\def\pasj{PASJ}
\def\pal{PASJ Letters}
\def\apj{ApJ}
\def\apjl{ApJ Letters}

\def\aj{AJ}
\def\aa{A\&A}

\def\nat{Nat}

\def\so{Sofue, Y.}
\def\ha{Handa, T.}
\def\na{Nakai, N.}
\def\fu{Fujimoto, M.}
\def\wa{Wakamatsu, K.}

\cen{\bf Ram-Pressure Stripping of Gas from Companions and}
\cen{\bf  Accretion onto a Spiral Galaxy: A Gaseous Merger}
\v\v
\cen{Yoshiaki SOFUE}

\cen{\it Institute of Astronomy, University of Tokyo, Mitaka, Tokyo 181, Japan}

\v\v\v\centerline{\bf Abstract} \v

We simulated the behavior of interstellar gas clouds in a companion galaxy
during a gas-dynamical interaction with the halo and disk of a spiral galaxy.
By ram pressure, the gas clouds are stripped from the companion, and accreted
toward the disk of the spiral galaxy.
If the companion's orbit is retrograde with respect to the rotation of the
spiral galaxy, infalling clouds hit the nuclear region.
Angular momentum transfer causes disruption of the inner gaseous disk, and
makes a void of interstellar gas in the bulge.
If the companion's orbit is either prograde or polar, infalling clouds are
accreted by the outer disk, and form a rotating gas ring.
We show that the ram-pressure stripping-and-accretion is one way from the
companion to a gas-rich larger galaxy, which causes disposal of interstellar
gas from the companion and effectively changes its galaxy type into earlier
(redder).
The ram-pressure process is significant durig merger of galaxies, in which
interstellar gas is stripped and accreted prior to the stellar-body merger.

Based on the simulation, we discuss a possible history of interstellar gas in
M31 system, which comprises M32, NGC 205 and possible merged galaxies.
The ram-pressure stripping explains the disposal of mass lost from evolving
stars in the dwarf elliptical companion M32.
The peculiar ``face-on'' spirals of ionized gas and dark clouds observed in
M31's bulge can be reproduced by a spiral inflow of accreting gas from the
companions and/or from the merger galaxy.

\v
{\bf Subject headings}: accretion --  dwarf ellipticals -- galaxies: individual
(M31, M32, NGC 205) -- interstellar matter -- merger -- ram-pressure.

\sect{\bf 1. Introduction} \v

Ram-pressure accretion of intergalactic clouds by galaxies with extended
gaseous halos has been studied by numerical simulations, and it was shown that
gaseous debris from tidally disrupted galaxies are soon accreted by an
encountering galaxy (Sofue and Wakamatsu 1991, 1992, 1993).
Observations of a considerable number of early-type galaxies, where gas is
moving perpendicularly or in a counter-rotating sense to the stellar rotation,
have suggested that the acquistion of gas from outside is not a rare event, and
sources of such cool  gas may be either tidal debris of disturbed galaxies or
intergalactic clouds (Sofue and Wakamatsu 1993).
On the other hand, ram-pressure stripping of interstellar matter by
intergalactic gas has been studied in relation to the origin of S0 galaxies
(e.g.,  Farouki and Shapiro 1980).

Tidal interaction beween galaxies has been extensively simulated numerically by
test-particle method (e.g., Toomre and Toomre 1972).
Even such a gaseous phenomenon as the Magellanic Stream (Mathewson et al. 1979)
has been modeled by a tidal debris from the Magellanic Clouds based on a
gravitational test-particle simulation (Fujimoto and Sofue 1976, 1977; Murai
and Fujimoto 1980).
Merging of galaxies has been studied by taking into account the dynamical
friction (Tremain 1976; Byrd 1979; Murai and Fujimoto 1980), and recent
$N$-body simulations of selfgravitating bodies showed more details of the
merging process among galaxies (e.g., Barnes 1989).
In these simulations, however, little attention has been paid to the gaseous
constituents, which must behave quite differently from stars for its viscous
and non-collisionless characteristics.

The giant Sb galaxy M31 and its companions (M32 and NGC 205) provide us a
unique opportunity to investigate tidal interaction among galaxies.
M32 is an anomalously compact elliptical, known for its tidal cut off and steep
central condensation, which is likely due to a tidal interaction with the M31's
disk (Kormendy 1987; Nieto and Prugniel 1987).
NGC 205 is a dwarf elliptical with less concentration and is known for its
gaseous content and star forming activity near the center (Johnson and
Gottesman 1983; Price and Grasdalen 1983; Bica et al 1990).
Tidal interaction of the companions with M31 has been numerically simulated
(Byrd 1976, 1977 1978; Sato and Sawa 1986), but no attention has been paid to
gaseous component and its hydrodynamical behavior.
Recent observations by the Hubble Space Telescope discovered a multiple nucleus
in M31, which gives definite proof of the merger of substantial galaxies with
M31 (Lauer et al. 1993). 
The central region of M31 exhibits a void of neutral as well as molecular
hydrogen gases (e.g., Brinks and Shane 1984; Koper 1990): M31 has no nuclear
gas disk, not alike the Milky Way.
On the other hand, M31' bulge exhibits a peculiar ``face-on'' spiral feature,
which is  observed both in the \halpha\ line emission (Ciardullo et al. 1988)
and in optical dark clouds (Sofue et al. 1993), although the gaseous mass is
much smaller comared to that found in our Galaxy.
This suggests that the inner gaseous disk of M31 has been disrupted by some
reason.

In this paper we simulate the stripping of interstellar matter from a companion
galaxy and its accretion onto a disk galaxy by applying the ram-pressure
accretion theory presented by Sofue and Wakamatsu (1991, 1992, 1993).
Based on results of simulations, we discuss the M31 system, and attempt to link
the properties observed in the dwarf elliptical companions as well as the
suggested merger galaxy to the peculiar morphology of gases observed in the
central region of M31.

\sect{\bf 2. Ram-Pressure Accretion Model and Basic Assumption}\v

\se{\it 2.1. HI and Molecular Clouds}\v

In this work, we treat ballistic orbits of interstellar gas clouds such as
giant molecluar clouds, which are initially distributed in companion galaxies
with random motion.
We assume that individual clouds are gravitationally bound.
HI clouds are easily stripped, but they can remain as clouds only when they are
massive enough, while less massive clouds are dissipated into the intergalactic
space.
In this paper, HI clouds are assumed to have  dimensions massive enough to be
gravitationally bound as the following:
radius $R \sim 500 $ pc;
density $\rho_{\rm HI}\sim 1$ $m_{\rm H}$ cm$^{-3}$;
 and mass  $m_{\rm HI}\sim 3.05 \times10^6 \Msun$.
The HI clould is gravitationally balanced with its internal motion (or rotation
) of $\sigma_{\rm HI}=5.11$ \kms.
For a molecular cloud, we assume a size and mass of the same order as those of
typical giant molecular clouds in our Galaxy:
radius $R \sim 30 $ pc;
mean density $\rho_{\rm cloud}\sim 10^2$ \htwo cm$^{-3}$; and
mass $m_{\rm cloud}\sim 1.32\times10^5 \Msun$.
The cloud is assumed to be gravitionally balancing with its internal velocity
dispersion of $\sigma_{\rm cloud}=4.36$ \kms.
Alternatively, the cloud may be maintained by rotation of the same amount of
velocity.

Although the cloud properties would vary due to interaction with the halo and
intergalactic gas, we here simply assume that their original properties are
kept during simulation.
We only comment the following:
the internal motion of clouds, such as turbulence, would be dissipated, which
would cause collapse of the cloud.
However, instabilities on the cloud surface due to interaction with the
intergalactic gas, such as the  Kelvin-Helmholtz instability, would excite and
to act maintain the internal motion.

\se{\it 2.2. Equations of Motion and Potentials of the Galaxies}\v

We adopt a simple ballistic model, as it has been adopted by Sofue and
Wakamatsu (1993) and  Farouki and Shapiro (1980).
The ram pressure (dynamical pressure) force on an intergalactic gas cloud (test
cloud) is given by $-\pi R^2 \rho({\bf r}) \Delta v^2$, where ${\bf \Delta
v}={\bf v - V}$ is the relative velocity of a cloud with respect to the halo
gas which is rotating at a velocity ${\bf V}$ in the potential of the disk
galaxy, and  $\rho({\bf r})$ is the density of diffuse  gas distributed around
the galaxy, which includes the intergalactic diffuse gas and the halo gas of
the galaxy.
The equation of motion for a test cloud which was initially distributed in the
companion galaxy can be written as;
$$
{{d^2{\bf r}}\over{dt^2}}
= \sum_{j=1}^2{{\partial \Phi_j}\over{\partial {\bf r}}}
- {{3 \rho({\bf r})}\over{4 R\rho_{\rm c}}} \Delta v {\bf \Delta v}, \eqno(1)
$$
where ${\bf v} = d{\bf r}/dt$ and  ${\bf r}=(x, y, z)$ are the cloud's velocity
and position with respect to the center of the disk galaxy,
 and $\rho_{\rm c}$ is the density of the cloud.
Here $\Phi_i$ are the gravitational potentials for the major galaxy ($j=1$) and
the companion galaxy ($j=2$).
The gravitational potential of the major galaxy is approximated by a modified
Miyamoto and Nagai's (1975) potential:
$$
\Phi_1= \sum_{i=1}^3  {GM_i \over \sqrt{ \varpi^2 + \left(a_i +
\sqrt{z^2+b_i^2}\right)^2 }}, \eqno(2)
$$
where $\varpi=(x^2+y^2)^{1/2}$,  $M_i$, $a_i$ and $b_i$ are the masses and
scale radii for the i-th mass component of the galaxy.
For the disk galaxy we assume three mass components: a central bulge ($i=1$);
disk ($i=2$); and a massive halo ($i=3$).
The parameters are given in Table 1.

For the companion galaxy we assume one component, for which $a_1=0$, so that
the potential represents a Plummer's law:
$$
\Phi_2 = {GM_{\rm C} \over {\sqrt{r^2 + b_{\rm C}^2}}}. \eqno(3)
$$
We assume that the mass of companions is an order of magnitude smaller than the
main body of the disk galaxy: $M_{\rm C}=0.1 M_2$, and that the center of mass
of the system coincides with the center of the major galaxy which is fixed to
the origin of the coordinates.

The motion of the companion galaxy is approximated by motion of a test particle
rotating around the galaxy in the potential described by equation (2).

The equation of motion of the center of a companion galaxy is written as
$$
{{d^2{\bf r}}\over{dt^2}}
= {{\partial \Phi_1}\over{\partial {\bf r}}}
- k M_{\rm C} \rho_{\rm MH}(r) {{\bf v}\over v}.
\eqno(4)
$$
The second term represents the dynamical friction due to the massive halo,
which is assumed to be at rest.
The density of the massive halo is assumed to be inversely proportional to the
square of $r$: $\rho_{\rm MH}(r)=\rho_{\rm MH0}(r/100~{\rm kpc})^{-2}$, with
$\rho_{\rm MH0}$ being a constant.
The variable  $k$ represents the coefficent of the dynamical friction.
Although coefficient $k$ is actually a slowly varying function of velocity and
mass (Tremain 1976; Byrd 1979),  we here assume it to be constant.
We took a value for $k \rho_{\rm MH0}$ so that the acceleration by the second
term becomes equal to 0.005 of the gravitational acceleration by the first term
 when the companion galaxy is at a distance of 100 kpc from the center.

\v \ce {-- Table 1--}\v

\se{\it 2.3. Models for Gaseous Disk, Halo and Intergalactic Gas}\v

We assume  a  density  distribution of a gaseous halo around the disk galaxy as
represented by the following equation:
$$
\rho({\rm r})
=\rho(\varpi, z)
= \rho_0
+{\rho_{\rm H} \over{(\varpi/\varpi_{\rm H})^2 + (z/z_{\rm H})^2 + 1 }}
+{\rho_{\rm D} \over{(\varpi/\varpi_{\rm D})^2 + (z/z_{\rm D})^2 + 1 }}
, \eqno(5)
$$
where $\rho_{\rm H}$, $\varpi_{\rm H}$ and $z_{\rm H}$ are parameters
representing the distribution of halo gas density,
$\rho_{\rm D}$, $\varpi_{\rm D}$ and $z_{\rm D}$ are those for the disk
component,
and $\rho_0$ is the  intergalactic gas density (Sofue and Wakamatsu 1993).
Values of the parameters are given in Table 2.

\cen{--Table 2--}

Since little is known about the rotation of halo gas, we here assume for
convenience that the halo gas is rotating around the $z$ axis with its
centrifugal force  balancing  the galaxy's gravity toward the $z$-axis:
$$
V(\varpi, z)=\sqrt{\sum_{i=1}^3 {GM_i\over \mu_i^3}} ~~ \varpi. \eqno(6)
$$
We assume that the gas is in a hydrostatic equilibrium in the $z$-direction, so
that $V_z=0$, and pressure gradient in $\varpi$ direction is neglected.
This assumption would be too simplified and the neglection of pressure gradient
in $\varpi$ direction may cause an overestimation of the rotation speed of the
halo gas.

\sect{\it 2.5. Initial Conditions}

We solve the differential equations by using the  Runge-Kutta-Gill method.
The time step of integration was taken to be smaller than 0.01 times the
dynamical time scale of each test particle (cloud) at the closest approach  to
the galaxy center.
Two cases of initial point of the companions are presented:  $(x,y,z)=(0, -100,
100)$ kpc and $(0, -50, 50)$ kpc.
Namely, the companions are put at sufficiently distant positios at a latitude
of 45\deg from the galactic plane.
Various different initial velocities are given to the companions, usually in
the sense that the initial position becomes the apogalactic point.
Fig. 1 illustrate the coordinate system and the orbit of companion.

\cen{ -- Fig. 1 -- }

$N$ interstellar clouds are initially distributed at random in the companion
within a radius $R$ and velocity dispersion $\sigma_v$, so that the ensemble of
test clouds are maintained to be a spherical system.
This might be replaced with a rotating disk of a similar size.
However, the initial distribuiton within the companion little affect the
accretion process after stripping, and, therefore, we adopt a spherical
distribution.

\sect{\bf 3. Simulation and Results}\v

\se{\it 3.1. Retrograde Encounter and  Polar-Spiral Accretion onto the
Nucleus}\v

Fig. 2 shows a result for a 45\deg\-inclined retrograde orbit, when the
integration started from the apogalacticon at $(x,y,z)=(0, -100, 100)$ kpc at a
tangential velocity of $(v_x, v_y, v_z)$
$=(-100, 0, 0)$ \kms.
The upper and lower panels show the $(x,z)$ and $(x,y)$ planes, respectively.
The spiral galaxy is rotating counter-clockwise in the $(x,y)$ plane.

The companion galaxy rotates around the disk galaxy on a semi-elliptical orbit.
Decrease in the apogalactic distance of the companion due to dynamical friction
is very gradual, and is almost negligible for the present simulation.
Orbits of individual gas clouds, which were distributed in the companion,
change drastically when they cross the galactic plane, where the clouds suffer
from strong ram-braking due to an almost head-on collision with the rotating
disk and halo gases.
The HI clouds are almost completely stripped from the companion, and fall
toward the disk of the major galaxy along polar orbits.
It is conspicuous that all stripped clouds are finally infall toward the
central region, where they form a polar ``accretion'' spiral.
The clouds are then accreted to the nuclear disk, and form a compact disk  with
a radius of a few kpc.

Stripping of molecular clouds occurs more gently, and many clouds survive the
stripping.
Accretion of stripped molecular clouds is also slower, and they remain as
intergalactic or intra-halo gaseous debris for a longer time than HI clouds,
but finally they are accreted by the galaxy.

Fig. 3a is the same as Fig. 2 but for a slower tangential velocity: $v_x=-80$
\kms.
As the perigalactic distance becomes smaller, stronger stripping of HI and also
of molecular clouds occurs.
Fig. 3b is the same but for a ``final'' stage of accretion of the molecular
clouds after 8 billion years, about two and half orbital rotations:
Although the accretion of molecular clouds are slow, they finally infall toward
the central region, except for some clouds which survive the stripping and
remain in the companion.

Fig. 4a and c are the same, but for $v_x=-70$ \kms, which might simulate M32.
Fig. 4b is an enlargement of Fig. 4a, which displays the central accretion in
more detail.
As the encounter with the disk is more direct in the sense of head-on
collision, the stripping occurs more drasticaly.
The stripped HI clouds infall toward the galaxy center along the accretion
spiral, and hit the nuclear region.
It is interesting to examine the tidal effect on the companion's stellar body.
Fig. 4d shows the result of test-particle simulation for the same condition as
the cloud but without ram pressure.
The star distribution is tidally deformed when the companion passes the
perigalactic point, but the effect is significant only in the outer part.

Fig. 5 is a case with a smaller distance of the initial position at $(x, y,
z)=(0, -50, 50)$ kpc with $(v_x, v_y, v_z)=(-50, 0, -50)$ \kms.
The result is similar to that for Fig. 4.

\ce{- Fig. 2, 3, 4, 5 -}

Similar results have been obtained for a wide range of initical conditions of
retrograde orbits.
Hence, we conclude that, if the companion galaxy approaches the major galaxy
on a retrograde orbit, the interstellar gas clouds are stripped, accreted to
the central region, and hit the nucleus, within about one revolution period, or
a couple of billion years.
Such an accretion of gas clouds fuels the nuclear region, and may explain the
peculiar polar spirals of ionized and molecular gases in the bulge of M31
(Ciardullo et al 1988; Sofue et al. 1993)(see section 4).
It may also have happened that the accretion in the nuclear region triggered
star burst in the center of the disk (Sofue and Wakamatsu 1993).

\se{\it 3.2.  Polar Encounter and Ring Formation}\v

Fig. 6 is a case for an almost polar (over-head) orbit starting from the same
initial position as Fig. 1 at $(x, y, z)$
$ = (0, -100, 100)$ kpc, but for $(v_x, v_y, v_z)$
$ = (-20, 50, 50)$ \kms, still in a retrograde sense.
The stripping occurs when the companion crosses the galactic plane, and
accreting clouds are merged by the rotating disk gas, forming a rotating ring
of about 10 kpc.

Fig. 7a shows a case of a polar encounter for an initial condition, $(x, y, z)$
$ = (0, -100, 100)$ kpc and $(v_x, v_y, v_z) $
$ = (0, 80, 0)$ \kms.
Fig. 7b is the result for  $(v_x, v_y, v_z)$
$ = (0, 0, -80)$ \kms, which might simulate NGC 205.
Stripping of HI clouds and their merging with the disk are similar to that
shown in Fig. 6,  and a ring of radius 10 kpc forms.
About half of the molecular clouds are stripped, but another half survive the
stripping.
Such formation of a gaseous ring may be related to the 10-kpc ring of HI and
molceular gases in M31 disk (see section 4.).

\cen{- Fig. 6, 7 -}

\se{\it 3.3. Prograde Encounter and Outer-Ring Formation}\v

Fig. 8 is a case of prograde (direct) orbit with the initial position and
velocity of
 $(x, y, z)=(0, -100, 100)$ kpc and $(v_x, v_y, v_z)=(+40, 0, 0)$ \kms.
The  stripping is rather mild, because the relative velocity of clouds against
the rotating disk and halo is smaller than that in the case of a retrograde
encounter.
The HI clouds are stripped during the close passage of the galactic plane, and
they begin to co-rotate with the disk gas by the friction, and finally attain a
ring of radius 15 kpc.
In such a prograde accretion, a part of the orbital angular momentum of the
infalling clouds will be transferred to that of disk rotation, and increases
the angular momentum of the disk, and therefore, increases the disk radius
(section 4).

Fig. 9 is the same as Fig. 8, but with larger tangential velocity, $v_x=+50$
\kms. Again a mild stripping of HI clouds and merging by the disk result in the
formation of large-radius ring.
It is also interesting to observe that outer spiral arms appear during the
accretion, which are warped from the disk plane.
On the other hand, only outer parts of molecular clouds of the companion are
stripped, but inner clouds remain unstripped.

Fig. 10 is the case for a closer orbit starting at $(0, -50, 50)$ kpc with
$(20, 50, 0)$ \kms.
A similar result is observed that ring of about 15 kpc radius forms.

\ce{- Fig. 8, 9, 10 - }

\se{\bf 4. Gaseous Merger in M31 System}\v

\sect{\it 4. 1. Orbits and Masses of M32 and NGC 205}\v

{\it M32}: Although some suggestions have been made (Byrd 1976, 1977; Sato and
Sawa 1986; Cepa and Beckman 1988), no definite orbits for the companions have
been determined, as was done for the Magellanic Clouds (Fujimoto and Sofue
1976, 1977).
 The projected distance of the present position of M32 from M31's center is
$21'.4=4.2$ kpc to the SW.
The measured distance and radial velocity of M31 are 670 kpc and $-299$ \kms,
respectively, and those of M32 are 660 kpc and $-217$ \kms (Allen 1973; de
Vaucouleurs et al. 1991).
Ford et al. (1978) found M32 to lie in front of M31 from observations of
planetary nebulae.
{}From these, we can estimate that M32 is about 10 kpc nearer to us and the
galacto-centric distance of about 11 kpc.
This is consistent with the difference in distances to M32 and M31-nucleus of
7.5 kpc as concluded by Byrd (1976).

M32 is approaching toward M31 at a relative line-of-sight velocity of +82 \kms.
The disk of M31 is rotating at a velocity of 220 \kms in the sense that the SW
side, where projected M32 is located, is approaching us.
Therefore, the line-of-sight velocity of M32 is retrograde with respect to
M31's rotation.
{}From the geometry of the two galaxies, the probability of retrograde orbit is
63\%.
We may thus suppose that M32 is rotating around M31 in a retrograde sense.
In this case M32's orbit and stripping-and-accretion process would be simulated
by results in Fig. 4 and 5.
Cepa and Beckman (1988) also suggested that M32' orbit is retrograde, while
they assumed that M32 is located farther than M31 and its present position is
almost in the disk of M31, which appears to be inconsistent with the
observations by Ford et al (1978).
Byrd (1976) took an opposite rotation based on a tidal warping model of the HI
disk, while parameters like the mass of M32 was taken much larger than the
recent value, so that the orbit determination must be revisited.

Obviously, the present position of M32 is too close to M31 for an initial
condition in our simulation, because our interest is in the stripping process
of their interstellar clouds, which we found to have occurred far before the
orbit shrinked to the present one by  the dynamical friction.
Hence, we may apply the present ram-pressure stripping-and-accretion model to
the past history of M32 several billions ago, when its galactocentric distance
was several tens of kpc.

{\it NGC 205}: The distance and radial velocity of NGC 205 are 640 kpc and
$-239$ \kms, respectively, or the galaxy is at  30 kpc nearer position to us
and is approaching M31 at a velocity of +60 \kms (see the literature above).
Since NGC 205 lies in the plane defined by the line of sight and the minor axis
[$(y, z)$ plane], the probabilities of retrograde and prograde are the same.
Here, we assume a polar orbit, as it was assumed by Sato and Sawa (1986), who,
however, put NGC 205 farther than M31.
Cepa and Beckman (1988) also suggested a high-inclination retrograde orbit for
NGC 205.
Fig 6, 7 and 8 would simlulate the case of such a polar-orbit case.

In both cases of M32 and NGC 205, however, their actual orbits around M31 are
still far from conclusive.
We also mention about the masses of the galaxies: the mass of M32 is around
$5\times10^8\Msun$ (Ford 1978; Byrd 1979), and NGC 205 around $10^{9 \sim
10}\Msun$ (Sato and Sawa 1986), which are both smaller than the mass assumed
for the companion used in the present simulation.
However, if we consider that the companions have experienced tidal truncation
of their outer maas during the past interactions with M31,
it will be not unreasonable to assume larger masses in the present simulation
where they are assumed to still contain substantial amount of interstellar gas.
Nevertheless, we should take the present result of simulation only as generic
prediction what happend to gas clouds in these galaxies.

\sect{\it 4.2. Galaxy Merger in M31?}\v

Recently, the Hubble Space Telescope observations discovered a mutiple nucleus
in M31, which gives definite proof of the merger of substantial galaxies with
M31 (Lauer et al. 1993). 
This fact implies that the gaseous stripping-and-accretion from companion
galaxies took place not only from M32 and NGC 205, but also from another merged
galaxy, which is now only seen as the second nucleus.
Since it is also likely that the merger galaxy contained insterstellar gas, the
present model applies also to this merger galaxy:
The ram-pressure stripping must have occured when the merging galaxy rotated on
an outer orbit, prior to the merging of the stellar body.
The stripped gas will have been accreted by M31's disk along a accretion
spiral, as simulated by the present models, and it must have behaved quite
differently from the merger body's motion.
Hence, the arguments given in the following subsections for the peculiar
gaseous features in M31's bulge applies also to gaseous remnant during the
merger.

\sect{\it 4.3. ``Face-on Spirals'' of Ionized and Molecular Gases in M31's
Center} \v

As the simulation for retrograde encounter indicates, accreting clouds that hit
the nuclear region exhibit peculiar kinematics, distinguished from the disk
rotation, and this may explain some  peculiar motions observed in the inner
region of M31.
CCD imaging of the \halpha\ emission of M31's bulge revealed peculiar spiral
patterns of ionized gas, and indicates an off-plane spirals which are rather
``face-on"  and exhibits anomalous velocities (Fig. 11: Ciardullo et al. 1988).
By image processing and color excess analyses of $B, ~V, ~R$, and $I$-band CCD
images of the central region of M31, we also found a face-on spiral feature of
dark clouds, which shows an excellent correlation with the \halpha\ spirals
(Fig. 12: Sofue et al. 1993).
Although the central 1-kpc region exhibits such peculiar and prominent
interstellar morphology, the gaseous mass itself is anomalously small (see
4.4).

\cen{--Fig. 11, 12 --}

The spiral pattern as defined by the \halpha\ emission and  dark clouds is
strongly asymmetric with respect to the galaxy center, and appears one-armed
and face-on.
This face-on spiral has no connection with the major disk and spiral arms of
M31 which are nearly edge-on at an inclination angle of 77\deg.
In Fig. 4c we showed the result of $N$-cloud simulation for the central region
for the retrograde encouter.
Except for scaling, the present simulation appears to nicely reproduce the
observed characteristics of the peculiar spiral features in Fig. 11 and 12.
In this model, the dark clouds in the spiral are considered to be stripped
clouds from the companion and/or merger galaxy, and the infalling motion and
friction among the clouds may have heated them to be ionized so that they are
observed in \halpha.

Here, we compared our result with the observations only qualitatively.
In order to obtain a better fitting to observations, which reveal much narrower
filamentary structures in the inner 1 kpc region, we need to investigate a more
detailed accretion process properly simulating hydrodynamical behavior of
individual clouds.
Namely, we need to adopt a more realistic density distribution of the gas disk,
and take into account variation of cloud properties during interaction with the
disk gas, as well as a change in the disk gas itself.

Stripping and accretion occurs most efficiently, when the companion crosses the
galactic plane.
Since the companions already rotated around M31 for several orbital periods,
they encoutered the M31 disk several times, and, accordingly, the
stripping-and-accretion must have occurred recurrently for several times.
The simulation indicates (Fig. 2 - 10) that molecular clouds infall more slowly
than HI clouds do, because of their weaker ram braking, and, hence, molecular
clouds can survive the accretion for more orbital rotations.
It is, therefore, also possible that the observed peculiar features in M31'
bulge are due to accretion of molecuar clouds which were stripped some
rotations ago.
This will apply particularly for a merger in which the dynmical time scale is
much shorter, and it happens that stripped clouds, particularly molecular
clouds, infall the nuclear region after the merger finished.

\sect{\it 4.4. Disruption of the Nuclear Gaseous Disk}

We have assumed that the disk gas in M31 has angular momentum large enough
compared to that of infalling gas clouds.
However, if we take into account the finite amount of angular momentum, the
rotation, particularly that of the inner region, of the disk gas must be
significantly affected by the accreting clouds (Sofue and Wakamatsu 1993).
If the encounter is retrograde, the  gas disk loses substantial amount of
angular momentum, and, as the consequence, the gaseous disk shrinks and results
in supplying the interstellar gas toward the nucleus.
This process might have caused a sudden increase of dense gas in the nuclear
region and triggered starburst (Sofue and Wakamatsu 1991).
The starburst would further cause fast outflow of interstellar gas due to
energy injection by sudden enhancment of supernova explosions and stellar
winds, and will result in disposal of gas from M31 bulge (e.g., Nakai et al
1987).

It will also happen that the vector of angular momentum changes significantly
by the accretion of external gas clouds, and the inner disk becomes tilted from
the original galactic plane.
In fact, the central 1 kpc of M31's bulge exhibits almost face-on spiral arms
as seen in \halpha\ and in dark clouds (Ciardullo et al. 1988; Sofue et al.
1993), which might be a mixture of the accreted gas and disk gas.

Hence, the inner gaseous disk is disrupted and dissipated by infalling external
gas clouds from the companions or the merger in such a way as either shrinking
toward the nucleus, expanding to become a ring, or becoming tilted from the
original plane.
If starburst occurs, it may also cause disposal of gas from the central region.
This scenario can well explain why the central region of M31 contains little
gas and shows an early-type characteristics.
In fact, observations show that the gaseous mass in the central 1kpc is
anomalously small, not exceeding some $10^7\Msun$ (Brinks and Shane 1984; Koper
et al 1990; Allen and Lequeux 1993; Sofue and Yoshida 1993), which is one or
two orders of magnitudes smaller compared to molecular gas mass in the central
region of normal Sb galaxies like the Milky Way and NGC 891 (Sofue and Nakai
1987, 1993).

\sect{\it 4.5. The 10-kpc Gaseous Ring}\v

If the encounter is prograde or polar, as is possibly the case for clouds from
NGC 205, clouds' orbital angular momentum is added to the disk angular
momentum, which results in expansion of the disk radius.
Accordingly, the inner disk gas is accumulated on a gasous ring at a larger
radius.
This will explain the 10-kpc ring of HI gas and a clear hole inside 8 kpc
(Brinks and Shane 1984).
This will also explain the large ring radius of molecular gas of about 10 kpc
(Koper et al. 1991), which is anomalously large compared to molecular-ring
radii of 4 - 5 kpc observed in usual Sb galaxies such as the Galaxy (Dame et al
1990) and NGC 891 (Sofue and Nakai 1987, 1993).

\sect{\it 4.6. Are M32 and NGC 205 Bulge Remnants of Disrupted Companions?}\v

The outer part of M32 is known to have been tidally  cut off, and it would have
lost a substantial fraction of the stellar disk (Nieto and Prugniel 1987).
This implies that M32 may have possessed a larger disk with more gas in the
past, but only the dense stellar bulge could have survived the tidal
disruption.
It would be, therefore, possible that M32 is a remnant of a bulge of a
disrupted companion.
This hypothesis explains the fact that M32 is anomalously compact and dense for
its small mass (Nieto and Prugniel 1987) and hosts a compact nucleus (Richstone
et al. 1990), not alkie usual dwarf ellipticals.
Moreover M32 has a surface-brightness distribution which is more similar to
that of a galactic bulge.

Another companion NGC 205 is known for its dark clouds (Sandage 1961) and
interstellar hydrogen (Johnson and Gottesman 1983) as well as for a recent
star-forming activity (Bica et al. 1983; Price and Grasdalen 1983).
As our simulation shows, it is easy for clouds in a companion to be stripped.
On the contrary, external clouds can hardly be captured by such a dwarf
elliptical as NGC 205 during its orbital motion around a large disk galaxy like
M31.
The existence of interstellar gas in NGC 205 can be explained only by a remnant
of interstellar gas which originally contained in the galaxy.
It is therefore also possible that NGC 205 is a remnant of a bulge of a
later-type galaxy that contained more amount of gas in the past and possiblly a
disk.

\sect{\bf 5. Discussion}

We have performed test-cloud simulations of ram-pressure
stripping-and-accretion of gas clouds from a companion galaxy onto its host
galaxy.
Obviously the result can be applied to systems of different scales by keeping
the similarity among mass, length, and time.
For example, without changing the masses, we can scale the length by a factor
of $f$ by scaling the time, velocity, and density  by factors of $f^{3/2}$,
$f^{-1/2}$, and $f^{-3}$, respectively.
Although the simulation gives a generic behavior of clouds in interacting
galaxy systems, we discussed particularly M31 and its companion as well as its
possible merger galaxy.

We discuss below more general problems which would be related to the present
model, and make some prediction based on this model.

\sect{\it 5.1. Disposal of Gas from Companions and  One-way Transfer}\v

Stripping-and-accretion of gas clouds from a companion to another galaxy should
play a substantial role in their evolution.
We stress that the gas transfer occurs much rapider than the stellar mass
merger due to dynamical friction.
The one-way transfer of gas results in a rapid evolution of the companion to
become redder (gas-poorer), and will lead to color contrast between the
companion (redder) and its host (bluer).
In fact, even at a glance of the Arp's (1966) atlas of peculiar galaxies, we
can find many multiple-galaxy systems in which larger galaxy is of later type
(spiral) and its companion is a dwarf elliptical or an early type such as a
small S0. A typical example is seen for M51 (Sc) and its companion (dwarf
elliptical irregular with a bar).

Ford (1978) showed that the gaseous mass observed in M32 is much less than that
expected from the mass loss rate from planetary nebulae, and stressed that the
disposal of mass lost from evolving stars in elliptical galaxies is an
outstanding problem.
An answer to this problem could be given by the present ram-pressure stripping
model:
such a small-mass companion as M32 ($\sim 5\times10^8\Msun$) would be severely
affected by ram stripping during its close encounter with the halo and disk of
M31, resulting in an almost entire disposal of interstellar gas.

\sect{\it 5.2. Fate of Gaseous Debris: Magellanic Stream}\v

The dynamical evolution of the triple system of the Galaxy, Large, and Small
Magellanic Clouds has been extensively studied by numerical simulations (e.g.,
Fujimoto and Sofue 1976, 1977).
In the current models of gravitaional interaction, the Magellanic Stream (MS)
of HI gas (Mathewson et al. 1979) has been reproduced as the tidal debris from
the SMC disturbed mainly by the LMC and partly by the Galaxy.
In these models, the HI debris were treated as test particles, and no
hydrodynamical effect has been taken into account.

However, if we take into account the ram-pressure effect on the MS, it will
rapidly infall toward the Galaxy, and will be accreted by the galactic disk
within about one Gyr.
Since the orbit of the MC is nearly polar, the accreting gas will be merged by
the rotating galactic disk at a radius of about 10 kpc.

We may convincingly conjecture that MS-like debris had been stripped rather
recurrently during the past tidal interactions among the three galaxies.
Hence, a greater number of MS-like clouds would be falling toward the Galaxy,
some which are already very close to the galactic plane or have merged.
Infalling directions are not necessarily related to the orbit of the companion,
but are polar spiral for a retrograde encounter, or they are in a
semi-corotation  with the galactic disk for a prograde encounter.
In either cse, however, they are approaching the galactic disk, and should be
observed to have negative velocities when looked at from the Sun.
Observed high-velocity HI clouds, which are accreting to the Galaxy at high
negative velocities (e.g., van Woerden et al. 1985), could be understood such
accreting clouds from companions.

\sect{\it 5.3. Fate of Gas in Merger}\v

Merger of galaxies has been numerically simulated using selfgravitating stellar
systems (Barnes 1989).
However, no particular analysis has been perfomred of the behavior of gaseous
constituents during the merger.
Since gas is not conllisionless and more viscous compared to stars in dynmical
interaction, gaseous constituents must interact more strongly.
As the simulation indicated, the gas stripping-and-accretion occurs much faster
than the dynamical merging time scale, even prior to merger starts.

However, if the perigalactic  distance is small enough and the companion
encounters the central bulge of the host galaxy, the merger time scale must be
much shorter.
In such a case, the gas stripping will occur even at a large distance before
the companion approaches the nucleus.
However, the accretion
of stripped clouds, particular molecular clouds,
are accreted in a longer, or a comparable, time scale than the merging time
scale.
This implies that gaseous accretion would continue even after the merger has
finished.
In either case, orbits of stripped gas clouds are quite different from that of
the stellar body.
The gaseous merger during  a galaxy merger will be subjected to a detailed
analsyis in a separate paper.

\v\v\v
The author thanks Dr. R. Ciardullo for providing him of the \halpha\ CCD image
of M31's bulge.
He is also indebted to the referee, Dr. G. G. Byrd, for valuable comments and
suggestions to the manuscript.

\v\v
\sect{\bf References}\v

\r Allen, C. W. 1973, Astrophysical Quantities (University of London, The
Athlone Press, London), Ch. 14.

\r Arp, H. C. 1966,  Atlas of Peculiar Galaxies (California Institute of
Technology, Pasadena).

\r Barnes, J. E. 1989, \nat, 338, 123.

\r Bica, E., Alloin, D., and Schmidt, A. A. 1990, \aa, 228, 23.

\r Brinks, E. and Shane, W. W.  1984,  \aa, 55, 179.

\r Byrd, G. G. 1976, \apj, 208, 688.

\r Byrd, G. G. 1977, \apj, 218, 86.

\r Byrd, G. G. 1978, \apj, 226, 70.

\r Byrd, G. G. 1979, \apj, 231, 32.

\r Cepa, J., and Beckman, J. E. 1988, \aa, 200, 21.

\r Ciardullo, R., Rubin, V. C.,Jacoby, G. H., Ford, H. C., Ford, Jr., W. K.
1988, \aj, 95, 438.

\r Dame, T. M., Ungerechts, H., Cohen, R. S., de Geus, E. J., Grenier, I. A.,
May, J., Murphy, D. C., Nyman, L.-A, and Thaddeus, P.  1987, \apj, 322, 706.

\r de Vaucouleurs G., de Vaucouleurs A., Corwin  H. G. Jr., et al.,
   1991, in Third Reference Catalogue of Bright Galaxies  (Springer Verlag, New
York).

\r Farouki, R., and Shapiro, L. 1980, \apj, 241, 928.

\r Ford, H. C. 1978, in Proc. IAU Symp. 76: Planetary Nebulae, Observations and
Theory (ed. Y. Tersian, D. Reidel Publ. Co), p.19.

\r Ford, H. C. Jacoby, G. H., and Jenner, D. C. 1978, \apj, 208, 683.

\r \fu, and \so\ 1976, \aa, 47, 263.

\r \fu, and \so\ 1977, \aa,  61, 199.

\r Johnson, D. W., and Gottesman, S. T. 1983, \apj, 275, 549.

\r Koper, E., Dame, T. M., Israel, F. P., Thaddeus, P. 1991, \apjl, 383, L11.

\r Kormendy, J. 1987, in Structure and Dynamics of Elliptical Galaxies (ed.
T.de Zeeuw, D. Reidel Publ. Co.), p.17.

\r Lauer, T., Faber, S. et al. 1993, \aj, October issure, in press.

\r Mathewson, D. S., Ford, V. L., Schwarz, M. P.,and Murray, J. D. 1979, in The
Large-scale Characteristics of the Galaxy, ed. W. B. Burton (Reidel Publishing
Co., Dordrecht), p. 547.

\r Miyamoto, M., and Nagai, R. 1975, \pa, {\bf 27}, 533.

\r Murai, T., and Fujimoto, M. 1980, \pasj, 32, 581.

\r \na, M. Hayashi, \ha, \so, T. Hasegawa, and M. Sasaki 1987, \pa, 39, 685.

\r Nieto, J. L. and Prugniel, P. 1987, in Structure and Dynamics of Elliptical
Galaxies (ed. T.de Zeeuw, D. Reidel Publ. Co.), p.99.

\r Price, J. S., and Grasdalen, G. L. 1983, \apj, 275, 559.

\r Richstone, D., Bower, G., and Dressler, A. 1990, \apj, 353, 118.

\r Sandage, A. 1961, in The Hubble Atlas of Galaxies, p.3.

\r Sato, N. R., and Sawa, T. 1986, \pa, 38, 63.

\r \so, \na, and \ha\ 1987, \pa, 39, 47.

\r \so, and \na\ 1993, \pa, 45, 139.

\r \so, and Yoshida, S. 1993, \apjl, in press.

\r \so, Yoshida, S., Aoki, T., Soyano, T., Tarusawa, K., and Wakamatsu, K.
1993, submitted to \pa.

\r \so, and \wa\ 1991, \pal, 43, L57.

\r \so, and \wa\ 1992, \pal,  44, L23.

\r \so, and \wa\ 1993, \aa, in press.

\r Toomre, A., and Toomre, J. 1972, \apj, 178, 632.

\r Tremaine, S. D. 1976, \apj, 203, 72.

\r van Woerden, H., Schwarz, U. J., and Hulsbosch, A. N. M. 1985, in The Milky
Way Galaxy, ed. H. van Woerden and R. J. Allen (Reidel Publishing Co.,
Dordrecht), p. 387.

\endpage

\settabs 6 \columns \v
\cen{Table 1: Parameters for gravitational potentials.}
\v\v
\hrule \vskip 0.5mm \hrule
\v\v
\+  ~~~ $ i $ &Mass component & & $M_i(\Msun)$ & $a_i$ (kpc) & $b_i$ (kpc) \cr
\v\v
\hrule
\v\v
\+ Disk galaxy$^\dagger$ \cr
\+ ~~~1   \dotfill  &Central bulge &  & $2.05\times 10^{10}$ & 0 & 0.495 \cr
\+ ~~~2 \dotfill  & Disk &  & $2.547 \times 10^{11}$ & 7.258 &  0.520 \cr
\+ ~~~3  \dotfill  & Massive halo & & $3 \times 10^{11}$ & 20 & 20 \cr
\v
\+ Companion$^\ddagger$  \dotfill  & Spheroid  &\hskip 20mm $M_{\rm C}=0.1~
M_2$ && 0 & 2 \cr
\v\v
\hrule
\v\v
\noi $\dagger$ Miyamoto-Nagai's (1975) potential with a modified massive halo.

\noi $\ddagger$ Plummer's potential. For a convinince, we adopted the same
parameter for M32 and NGC 205, which little affects the result of accretion
after stripping, while the fraction of unstripped molecular clouds would change
by different masses.

\endpage

\settabs 3 \columns \v
\cen{Table 2: Parameters for gaseous components.}
\v\v
\hrule \vskip 0.5mm \hrule
\v\v
\+ Intergalactic gas: & $\rho_0$ 	~\dotfill~ & $10^{-5}\MH$ \pcc \cr
\v
\+ Halo: & $\rho_{\rm H}$ 	~\dotfill~ &  0.01 \mH \pcc\cr
\+ 	& $\varpi_{\rm H}$ 	~\dotfill~ & 15 kpc \cr
\+ 	& $z_{\rm H}$ 	~\dotfill~ & 10 kpc \cr
\v
\+ Disk: & $\rho_{\rm D}$ 	~\dotfill~ & 1 \mH \pcc \cr
\+ 	& $\varpi_{\rm D}$ 	~\dotfill~ & 10 kpc \cr
\+ 	& $z_{\rm D}$ 	~\dotfill~ & 0.2 kpc \cr
\v
\+ Molecular cloud:  & $\rho_{\rm cloud}$ ~\dotfill~ & $100$ \htwo \pcc \cr
\+ 	& $R$	~\dotfill~ & 30 pc \cr
\+ 	& $m=(4\pi/3)R^3\rho_{\rm HI}$ ~\dotfill~ & $ 1.32\times10^5 \Msun$ \cr
\+ 	& $\sigma_{\rm cloud}$	~\dotfill~ & 4.36 \kms \cr
\v
\+ HI cloud:  & $\rho_{\rm HI}$ 	~\dotfill~ & $1$ \mH \pcc \cr
\+ 	& $R$	~\dotfill~ & 500 pc \cr
\+ 	& $m=(4\pi/3)R^3\rho_{\rm HI}$ ~\dotfill~ & $ 3.05\times10^6\Msun$  \cr
\+ 	& $\sigma_{\rm HI}$	~\dotfill~ & 5.11 \kms \cr
\v\v
\hrule

\page

\noi{Figure Captions} \vskip 3mm

\r Fig. 1: Coordinate system and orientation of the disk galaxy with its
rotation direction used in the simulation.
Possible orbits of M32 and NGC 205 as discussed in the text and the line of
sight are illustrated schematically.

\v
\r Fig. 2: Stripping of HI (asterisques) and molecular (dots) clouds from a
companion  and their ram-pressure accretion toward M31 for a semi-retrograde
orbit.
The inititial condition is $(x, y, z)=(0, -100, 100)$ kpc and $(v_x, v_y,
v_z)=(-100, 0, 0)$ \kms.
Upper panel is the projection onto the $(x, z)$ plane (M31 is edge on), which
mimick approximately the view when seen from us, and the lower panel onto
$(x,y)$ plane (M31's disk plane).
M31 is rotating unti-clockwise on the $x,y$ plane.
The cross indicates the center of M31, and the square represent the companion
plotted every 0.1 Gyr. Clouds are also plotted every 0.1 Gyr.

\v
\r Fig. 3a: The same as Fig.2 but for a prograde orbit of the companion.
The inititial condition is $(x, y, z)=(0, -100, 100)$ kpc and $(v_x, v_y,
v_z)=(-80, 0, 0)$ \kms.

\r Fig. 3b: The same but after 8 Gyr. Molecular clouds are also accreted toward
M31's center.

\v
\r Fig. 4a: The same as Fig.2, but for the inititial condition $(x, y, z)=(0,
-100, 100)$ kpc and $(v_x, v_y, v_z)=(-70, 0, 0)$ \kms.
This case simulate M32.

\r Fig. 4b: Enlargement of (a), in order to see the central accretion spirals.
The clouds are distributed spirally with a large tilt angle from the galactic
plane. This mimick the peculiar ``face-on" spiral feature as observed in
molecular clouds (Sofue et al. 1993) and in \halpha\ (Ciardullo et al. 1988) as
shown in Fig. 11 and 12.
Plot interval is 0.1 Gyr

\r Fig. 4c: The same as (a) but until 4 Gyr, plotted every 0.2 Gyr.

\r Fig. 4d. The same as Fig. 4c, but for the stellar component, for which only
the gravitation affect,  but no ram force.

\v\r Fig. 5: The same as Fig. 1, but for a closer orbit with
an inititial condition $(x, y, z)=(0, -50, 50)$ kpc and $(v_x, v_y, v_z)=(-50,
0, -50)$ \kms. Plot interval is 0.2 Gyr.
The display is a stereogram in order to obtain an impression of the
three-dimensional structure of the simulated feature.

\v\r Fig. 6: The same as Fig. 1, but for a semi-polar orbit with
an inititial condition $(x, y, z)=(0, -100, 100)$ kpc and $(v_x, v_y,
v_z)=(-20, 50, 50)$ \kms.
Stripped clouds form a ring of radius 10 kpc.
Plot interval is 0.2 Gyr.

\v\r Fig. 7a: Polar encounter for an
inititial condition  $(x, y, z)=(0, -100, 100)$ kpc and $(v_x, v_y, v_z)=(0,
80, 0)$ \kms.
Stripped clouds form a 10-kpc ring.
Plot interval is 0.2 Gyr.
The display is a stereogram.

\r Fig. 7b: Polar encounter for an
inititial condition  $(x, y, z)=(0, -100, 100)$ kpc and $(v_x, v_y, v_z)=(0, 0,
-80)$ \kms.
This mimicks the orbit of NGC 205.
The display is a stereogram.

\v\r Fig. 8: Prograde (direct) encouter for an
inititial condition  $(x, y, z)=(0, -100, 100)$ kpc and $(v_x, v_y, v_z)=(+20,
0, -60)$ \kms.
A ring is formed.

\v\r Fig. 9: Prograde encounter for an
inititial condition  $(x, y, z)=(0, -100, 100)$ kpc and $(v_x, v_y, v_z)=(+50,
0, 0)$ \kms.
A ring or larger radius is formed.
Also transient warped spiral arms appear in the outer region.
Plot interval is 0.2 Gyr.

\v\r Fig. 10: Prograde encounter for a closer orbit for an
inititial condition  $(x, y, z)$
$=(0, -50, 50)$ kpc and $(v_x, v_y, v_z)=(20, 50, 0)$ \kms.
A ring is formed at a radius of 15 kpc.

\v
\r Fig. 11: \halpha\ image of the central $2\times 2$ kpc region of M31
(reproduced from Ciardullo et al. 1988; courtesy of Dr. R. Ciardullo). A
face-on spiral structure is seen in a good coincidence with the dark cloud
feature in Fig. 12.
The feature is reproduced qualitatively by the simulation as shown in Fig. 4b.

\v
\r Fig. 12: $B-V$ image of the central 2.4 kpc$\times$2.4 kpc region of M31,
which was observed with the Kiso 105-cm Schmidt telescope (reproduced from
Sofue et al. 1993).
Dark clouds in the central 2 kpc are found to trace a face-on, one-armed spiral
structure, which has no apparent connection with the major disk of M31, but
comprises a disk perpendicular to M31's disk plane.
The feature can be explained by the accretion spiral as simulated by the model
shown in Fig. 4.

\bye